\documentclass[aps,prd,onecolumn,preprintnumbers,groupedaddress,showpacs,nofootinbib,amssymb]{revtex4}
\usepackage{graphicx,color}
\usepackage{amsmath}
\usepackage{amssymb}
\usepackage{amsfonts}
\usepackage{bm}
\usepackage{cancel}

\allowdisplaybreaks[4]

\begin{document}

\newcommand{\be}{\begin{equation}}
\newcommand{\ee}{\end{equation}}
\newcommand{\bea}{\begin{eqnarray}}
\newcommand{\eea}{\end{eqnarray}}
\newcommand{\nn}{\nonumber \\}
\newcommand{\e}{\mathrm{e}}
\newcommand{\tr}{\mathrm{tr}\,}

\newcommand{\br}{\bm{r}}

\title{ Newton law  in covariant unimodular $F(R)$ gravity}

\author{S.~Nojiri,$^{1,2}$\,\thanks{nojiri@gravity.phys.nagoya-u.ac.jp}
S.~D.~Odintsov,$^{3,4}$\,\thanks{odintsov@ieec.uab.es}
V.~K.~Oikonomou,$^{5,6}$\,\thanks{v.k.oikonomou1979@gmail.com}}

\affiliation{$^{1)}$ Department of Physics, Nagoya University, Nagoya
464-8602,
Japan \\
$^{2)}$ Kobayashi-Maskawa Institute for the Origin of Particles and the
Universe, Nagoya University, Nagoya 464-8602, Japan \\
$^{3)}$Institut de Ciencies de lEspai (IEEC-CSIC),
Campus UAB, Carrer de Can Magrans, s/n
08193 Cerdanyola del Valles, Barcelona, Spain \\
$^{4)}$ ICREA, Passeig LluAs Companys, 23,
08010 Barcelona, Spain \\
$^{5)}$ Tomsk State Pedagogical University, 634061 Tomsk, Russia \\
$^{6)}$ Laboratory for Theoretical Cosmology, Tomsk State University of
Control
Systems and Radioelectronics (TUSUR), 634050 Tomsk, Russia \\
}

\tolerance=5000

\begin{abstract}

We propose a covariant ghost-free unimodular $F(R)$ gravity theory, which contains a
three-form field and study its structure using the analogy of the proposed theory with a quantum
system which describes a charged particle in uniform magnetic field. Newton's law in non-covariant unimodular $F(R)$ gravity as well as in
unimodular Einstein gravity is derived and it is shown to be just the same as
in General Relativity. The derivation of Newton's law in covariant unimodular
$F(R)$ gravity shows that it is modified precisely in the same way as in
the ordinary $F(R)$ theory. We also demonstrate that the cosmology of a Friedmann-Robertson-Walker background, is equivalent in the non-covariant and covariant formulations of unimodular $F(R)$ theory.

\end{abstract}


\maketitle

\section{Introduction}

The (non-covariant) unimodular Einstein gravity is supposed to solve the
cosmological constant problem in a geometrical way (see for example
Ref.~\cite{Alvarez:2005iy} for an introduction).
At the same time, in some sense such a theory is less dynamical than
General Relativity with cosmological constant. Recently, the
unimodular $F(R)$ gravity has been proposed 
\cite{Nojiri:2015sfd},  which is quite rich dynamically
and successfully describes inflationary or bouncing cosmology.
However, its further development, refinement and better understanding of its physical
properties is quite desirable. The present paper is devoted to the construction of a covariant formulation
of unimodular $F(R)$ gravity, as well as to the study of the most important
physical implications, like study of Newton's law and the Friedmann-Robertson-Walker (FRW) cosmology of such
theory.

Our first aim with this letter, is to study Newton's law in unimodular $F(R)$ gravity, focusing on the possible modifications that the unimodular formalism may bring along. An exciting new result that we obtain is that Newton's law in unimodular $F(R)$ gravity does not receive any new corrections, so basically it is the same as in General Relativity. Also as a second task, we formulate a covariant unimodular $F(R)$ gravity theory and we analyze its dynamical structure by using the analogy of the covariant unimodular $F(R)$ theory with a simple quantum mechanical system of a charged particle in a uniform magnetic field. For the covariant unimodular $F(R)$ theory, we derive Newton's law and we demonstrate that it is modified in the same way as in ordinary $F(R)$ gravity, and this result is quite different from the non-covariant unimodular theory. We also demonstrate that the equations of motion for the covariant unimodular $F(R)$ theory for a FRW background are equivalent to the equations of motion corresponding to the non-covariant theory. Finally, we discuss possible generalizations of the covariant formulation.

This letter is organized as follows: In section II we discuss Newton's law in the context of unimodular $F(R)$ gravity, while in section III we introduce the covariant version of unimodular $F(R)$ gravity, and also we study Newton's law and its modifications due to the covariant theory. The concluding remarks with a brief critical discussion follow in the end of the paper.

\section{Newton law in the unimodular $F(R)$ gravity}

In this section we  study  Newton's law in unimodular $F(R)$ gravity, with the unimodular
constraint being equal to,
\be
\label{UF1}
\sqrt{ -g } = 1\, .
\ee
The constraint can be realized by using a Lagrange multiplier field
$\lambda$, so that the unimodular $F(R)$ gravity action is the following,
\be
\label{UF2}
S = \int d^4 x \left\{ \sqrt{-g} \left( \frac{F(R)}{2\kappa^2} - \lambda
\right) + \lambda \right\}
+ S_\mathrm{matter} \left( g_{\mu\nu}, \Psi \right)\, ,
\ee
where $S_\mathrm{matter}$ denotes action of the matter fluids present, and $\Psi$ denotes the
matter fluids. It is known that the above theory is ghost-free, as was discussed in Ref. \cite{Nojiri:2015sfd}. It can be generalized by multiplying second $\lambda$ function in Eq.~(\ref{UF2}) by
$\left( \sqrt{-g}\right)^q$, with $q$ being an arbitrary non-vanishing constant.
As in the standard $F(R)$ gravity theory, we may rewrite the action (\ref{UF2}) in the scalar-tensor form,
\be
\label{UF3}
S = \int d^4 x \left\{ \sqrt{-g} \left( \frac{1}{2\kappa^2} \left( R
 - \frac {3}{2} g^{\mu\nu} \partial_\mu \phi \partial_\nu \phi - V(\phi)
\right)
 - \lambda \e^{2\phi} \right) + \lambda \right\}
+ S_\mathrm{matter} \left( \e^\phi g_{\mu\nu}, \Psi \right)\, ,
\ee
where the potential $V(\phi)$ is given by,
\be
\label{UF4}
V(\phi) = \frac{A(\phi)}{F'\left( A \left( \phi \right) \right)}
 - \frac{F\left( A \left( \phi \right) \right)}{F'\left( A \left( \phi
\right)
\right)^2}\, ,
\ee
and the function $A(\phi)$ is defined by solving the equation $\phi = - \ln F'(A)$. The unimodular constraint (\ref{UF1}) is now modified to be
$\e^{2\phi}\sqrt{-g} = 1$. Eliminating the scalar field $\phi$, the action (\ref{UF3}) can be rewritten
as follows,
\be
\label{UF7}
S = \int d^4 x \sqrt{-g} \left( \frac{1}{2\kappa^2} \left( R - \frac{3}{32
g^2}
g^{\mu\nu} \partial_\mu g \partial_\nu g - V\left(\frac{1}{4}\ln \left( - g
\right) \right) \right) \right)
+ S_\mathrm{matter} \left( \left( -g \right)^\frac{1}{4} g_{\mu\nu}, \Psi
\right)\, .
\ee
We consider the perturbation of the metric $g_{\mu\nu}$ around
the background metric $g_{\mu\nu}^{(0)}$, as follows, $g_{\mu\nu} = g_{\mu\nu}^{(0)} + h_{\mu\nu}$.
By assuming that the background metric is flat, that is, $g_{\mu\nu}^{(0)} = \eta_{\mu\nu}$, we finally find that,
\be
\label{UF10}
\sqrt{- g} R \sim - \frac{1}{2} \partial_\lambda h_{\mu\nu} \partial^\lambda
h^{\mu\nu}
+ \partial_\lambda h^\lambda_{\ \mu} \partial_\nu h^{\mu\nu}
  - \partial_\mu h^{\mu\nu} \partial_\nu h + \frac{1}{2}\partial_\lambda h
\partial^\lambda h  \, ,
\ee
where $h$ is the trace of the tensor field $h_{\mu\nu}$, $h \equiv \eta^{\rho\sigma}
h_{\rho\sigma}$. Because of the flat background choice, we find $V(0)=V'(0)=0$,
and one may write down the potential $V$ as  $V \sim \frac{1}{2}m^2 h^2$. The linearized action has the following form
\begin{align}
\label{UF12}
S =& \frac{1}{2\kappa^2} \int d^4 x \left\{
 - \frac{1}{2} \partial_\lambda h_{\mu\nu} \partial^\lambda h^{\mu\nu}
+ \partial_\lambda h^\lambda_{\ \mu} \partial_\nu h^{\mu\nu}
 - \partial_\mu h^{\mu\nu} \partial_\nu h + \frac{1}{2}\partial_\lambda h
\partial^\lambda h
 - \frac{3}{32}\partial_\mu h \partial^\mu h - \frac{1}{2}m^2 h^2 \right\}
\nn
& + S_\mathrm{matter} \left( \eta_{\mu\nu} + h_{\mu\nu} -
\frac{1}{4}\eta_{\mu\nu} h , \Psi \right)\, .
\end{align}
Then by varying with respect to $h_{\mu\nu}$, we obtain the following equations,
\be
\label{UF13}
\partial_\lambda \partial^\lambda h_{\mu\nu}
 - \partial_\mu \partial^\lambda h_{\lambda\nu}
 - \partial_\nu \partial^\lambda h_{\lambda\mu}
+ \partial_\mu \partial_\nu h
+ \eta_{\mu\nu} \partial^\rho \partial^\sigma h_{\rho\sigma}
 - \frac{13}{16} \eta_{\mu\nu} \partial_\lambda \partial^\lambda h  - m^2
\eta_{\mu\nu} h
= \kappa^2 \left( T_{\mu\nu} - \frac{1}{4} \eta_{\mu\nu} T \right) \, ,
\ee
where $T_{\mu\nu}$ stands for the energy-momentum tensor of the matter fluids, and $T$ is the
trace of $T_{\mu\nu}$, $T \equiv \eta^{\rho\sigma} T_{\rho\sigma}$.
By multiplying Eq.~(\ref{UF13}) by $\eta^{\mu\nu}$, we obtain,
\be
\label{UF13b}
0 = - \frac{5}{4} \partial_\lambda \partial^\lambda h - 4 m^2 h
+ 2 \partial^\mu \partial^\nu h_{\mu\nu}\, .
\ee
In order to investigate Newton's law, we consider a point source at the
origin, so that the energy-momentum tensor has the following components, $T_{00} = M \delta \left( \br
\right)$,
$T_{ij} = 0$ $\left(i,j=1,2,3\right)$, and we look for a static solution of Eq.
(\ref{UF13}). The $(0,0)$, $(i,j)$, and $(0,i)$ components of Eq.~(\ref{UF13}) and
Eq.~(\ref{UF13b}) have the following form:
\begin{align}
\label{UF15}
& \partial_i \partial^i h_{00} - \partial^i \partial^j h_{ij} +
\frac{13}{16}
\partial_i \partial^i h
+ m^2 h = \frac{3 \kappa^2}{4} M \delta \left( \br \right) \, , \\
\label{UF16}
& \partial_k \partial^k h_{ij} - \partial_i \partial^k h_{kj} - \partial_j
\partial^k h_{ki}
+ \partial_i \partial_j h + \delta_{ij} \partial^k \partial^l h_{kl}
 - \frac{13}{16} \delta_{ij} \partial_k \partial^k h  - m^2 \delta_{ij} h
= \frac{\kappa^2}{4} M \delta \left( \br \right) \, , \\
\label{UF17}
& \partial_j \partial^j h_{0i} - \partial_i \partial^k h_{k0} = 0 \, ,\\
\label{UF18}
& - \frac{5}{4} \partial_k \partial^k h - 4 m^2 h + 2 \partial^i \partial^j
h_{ij} = 0 \, .
\end{align}
In the Einstein-Hilbert gravity case, there exist four gauge degrees of freedom, but in
case of the unimodular $F(R)$ gravity, there exist only three gauge degrees of
freedom, due to the unimodular constraint (\ref{UF1}). Then we now impose three gauge conditions, $\partial^i h_{ij} = 0$, in which case Eq.~(\ref{UF18}) reduces to,
\be
\label{UF20}
 - \frac{5}{4} \partial_k \partial^k h - 4 m^2 h = 0 \, ,
\ee
and under a proper boundary condition, we obtain $h=0$. By using the three gauge conditions $\partial^i h_{ij} = 0$ and the equation
$h=0$,
we can rewrite Eqs.~(\ref{UF15}), (\ref{UF16}), and (\ref{UF17}) as
follows,
\be
\label{UF22}
\partial_i \partial^i h_{00} = \frac{3 \kappa^2}{4} M \delta \left( \br
\right)
\, , \quad
\partial_k \partial^k h_{ij} = \frac{\kappa^2}{4} M \delta \left( \br
\right)
\, , \quad  \partial_j \partial^j h_{0i} - \partial_i \partial^k h_{k0}
= 0 \, .
\ee
Under a proper boundary condition, the above equations and the equation
$h=0$, yield the following,
\be
\label{UF25}
h_{0i}=0\, , \quad h_{ij} = \frac{1}{3} \delta_{ij} h_{00}\, .
\ee
Defining the Newtonian potential $\Phi$ by $h_{00} = 2 \Phi$, Eq.~(\ref{UF22})
gives the Poisson equation for the Newtonian potential $\Phi$,
$\partial_i \partial^i \Phi = \frac{3 \kappa^2}{8} M \delta
\left( \br \right)$.
Hence, by redefining the gravitational constant $\kappa$ by
$\frac{3 \kappa^2}{4} \to \kappa^2 = 8\pi G$,
we obtain the standard Poisson equation for the Newtonian potential $U$,
$\partial_i \partial^i \Phi = 4\pi G M \delta \left( \br \right)$,
whose solution is given by,
\be
\label{UF29}
\Phi = - \frac{GM}{r}\, .
\ee
The above result is quite different from the case of the standard $F(R)$
gravity (for review, see
\cite{Capozziello:2011et,Nojiri:2010wj,Nojiri:2006ri,Capozziello:2010zz}),
where the propagation of the scalar mode
$\phi = - \ln F'(A)$ gives a non-trivial correction to Newton's law of gravity.
In the case of unimodular $F(R)$ gravity, owing to the fact that the unimodular condition
(\ref{UF1}) can be rewritten as follows, $\e^{2\phi}\sqrt{-g} = 1$, the degree
of the freedom in the scalar mode $\phi$ is actually eliminated from the field equations, and therefore $\phi$ does not
propagate. Therefore, as a result no correction to Newton's law of gravity appears.

Let us compare the above situation with Newton's law of gravity in the context of unimodular
Einstein gravity, in which case $F(R)=R$. In the unimodular Einstein gravity, there exists a solution describing the
flat
space-time, $g_{\mu\nu}=\eta_{\mu\nu}$, $\lambda=0$. By considering the perturbation $g_{\mu\nu} = \eta_{\mu\nu} + h_{\mu\nu}$,
we find that the linearized action has the following form,
\begin{align}
\label{UE5}
S =& \frac{1}{2\kappa^2} \int d^4 x \left\{   - \frac{1}{2} \partial_\lambda
h_{\mu\nu} \partial^\lambda h^{\mu\nu}
+ \partial_\lambda h^\lambda_{\ \mu} \partial_\nu h^{\mu\nu}
 - \partial_\mu h^{\mu\nu} \partial_\nu h + \frac{1}{2}\partial_\lambda h
\partial^\lambda h - \frac{1}{2} \lambda h \right\} \nn
& + S_\mathrm{matter} \left( \eta_{\mu\nu} + h_{\mu\nu} , \Psi \right)\, .
\end{align}
By varying with respect to $h_{\mu\nu}$, we obtain the following equations,
\be
\label{UE6}
\partial_\lambda \partial^\lambda h_{\mu\nu} - \partial_\mu \partial^\lambda
h_{\lambda\nu}
 - \partial_\nu \partial^\lambda h_{\lambda\mu} + \partial_\mu \partial_\nu
h
+ \eta_{\mu\nu} \partial^\rho \partial^\sigma h_{\rho\sigma} - \eta_{\mu\nu}
\partial_\lambda \partial^\lambda h - \frac{1}{2} \eta_{\mu\nu} \lambda
= \kappa^2 T_{\mu\nu} \, .
\ee
The unimodular constraint (\ref{UF1}), which can be obtained by the
variation of the action with respect to $\lambda$, has the form of $h=0$.
We now obtain $h=0$ by the unimodular constraint but in case of $F(R)$
gravity
in the scalar-tensor form (\ref{UF3}), the constraint $\e^{2\phi}\sqrt{-g} =
1$
is solved with respect to the scalar field $\phi$, therefore the metric itself is not constrained.
Multiplying Eq.~(\ref{UE6}) by $\eta^{\mu\nu}$ and using (\ref{UE7}), we obtain,
\be
\label{UE7}
 - 2 \partial_\lambda \partial^\lambda h + 2 \partial^\rho \partial^\sigma
h_{\rho\sigma} - 2 \lambda = \kappa^2 T\, .
\ee
By solving Eq.~(\ref{UE7}) with respect to $\lambda$ we get,
$\lambda
= - \partial_\lambda \partial^\lambda h + \partial^\rho \partial^\sigma
h_{\rho\sigma}
 - \frac{\kappa^2}{2} T$.
Then by deleting $\lambda$ and taking account of Eq. (\ref{UE7}), we may rewrite Eq.~(\ref{UE6}) as follows,
\be
\label{UE9}
\partial_\lambda \partial^\lambda h_{\mu\nu} - \partial_\mu \partial^\lambda
h_{\lambda\nu}
 - \partial_\nu \partial^\lambda h_{\lambda\mu} + \partial_\mu \partial_\nu
h
+ \frac{1}{2} \eta_{\mu\nu} \partial^\rho \partial^\sigma h_{\rho\sigma}
 - \frac{1}{2} \eta_{\mu\nu} \partial_\lambda \partial^\lambda h
= \kappa^2 \left( T_{\mu\nu} - \frac{1}{4} \eta_{\mu\nu} T \right) \, .
\ee
In order to investigate Newton's law of gravity, we consider a point source at the
origin, with the components of the energy-momentum tensor being, $T_{00} = M \delta \left( \br \right)$, $T_{ij} = 0$
$\left(i,j=1,2,3\right)$
and we look for a static solution of Eq. (\ref{UE9}).Then the $(0,0)$, $(i,j)$, and $(0,i)$ components of Eq.~(\ref{UE9}) are:
\begin{align}
\label{UE10}
& \partial_i \partial^i h_{00} - \frac{1}{2} \partial^i \partial^j h_{ij}
+ \frac{1}{2} \partial_i \partial^i h
= \frac{3 \kappa^2}{4} M \delta \left( \br \right) \, , \\
\label{UE11}
& \partial_k \partial^k h_{ij} - \partial_i \partial^k h_{kj} - \partial_j
\partial^k h_{ki} + \frac{1}{2} \delta_{ij} \partial^k \partial^l h_{kl}
 - \frac{1}{2} \delta_{ij} \partial_k \partial^k h
= \frac{\kappa^2}{4} M \delta \left( \br \right) \, , \\
\label{UE12}
& \partial_j \partial^j h_{0i} - \partial_i \partial^k h_{k0} = 0 \, .
\end{align}
By imposing the gauge conditions $\partial^i h_{ij} = 0$,
Eqs.~(\ref{UE10}), (\ref{UE11}), and (\ref{UE12}) reduce to
(\ref{UF22}), again.
Therefore by the redefinition of the gravitational constant $\kappa$,
$\frac{3 \kappa^2}{4} \to \kappa^2 = 8\pi G$,
we obtain Newtonian potential of Eq.  (\ref{UF29}).

Hence, although there are some differences between the unimodular $F(R)$
gravity and the unimodular Einstein gravity, especially in the degrees of
freedom, an identical Newtonian potential is produced. The differences in the number of degrees of freedom however, may affect the
cosmological perturbations, and therefore the structure formation in the Universe.

Let us now briefly review the FRW cosmology in the unimodular $F(R)$ gravity of Eq.
(\ref{UF2}) based on
\cite{Nojiri:2015sfd,Nojiri:2016ygo,Saez-Gomez:2016gum}.
In terms of the cosmological time $t$, the metric of the FRW universe
with a flat spatial part, does not satisfy the unimodular constraint
(\ref{UF1}).
This problem is solved by defining ``unimodular cosmological time'' $\tau$ instead
of
the standard cosmological time $t$, which satisfies $d\tau = a(t)^3 dt$.
By using the unimodular cosmological time $\tau$, the FRW metric can be rewritten in
the unimodular form:
\be
\label{UniFRW}
ds^2 = - a\left(t\left(\tau\right)\right)^{-6} d\tau^2 +
a\left(t\left(\tau\right)\right)^{2}
\sum_{i=1}^3 \left( dx^i \right)^2 \, ,
\ee
and hence the unimodular constraint is satisfied.
Using the unimodular metric of Eq.~(\ref{UniFRW}), and by making use of the
Lagrange multiplier method \cite{Lim:2010yk,Capozziello:2010uv},
the vacuum Jordan frame unimodular $F(R)$ gravity action is given in Eq.
(\ref{UF2}).
Then by varying with respect to the metric, we obtain the following
equations of motion,
\be
\label{Uni12}
0=\frac{1}{2}g_{\mu\nu} \left( F(R) - 2\kappa^2 \lambda \right) - R_{\mu\nu}
F'(R) + \nabla_\mu \nabla_\nu F'(R) - g_{\mu\nu}\nabla^2 F'(R)
+ \frac{\kappa^2}{2} T_{\mu\nu} \, .
\ee
and the FRW equations take the following form,
\begin{align}
\label{Uni14}
0 = & - \frac{a^{-6}}{2} \left( F(R) - 2\kappa^2 \lambda \right) + \left( 3
\dot K + 12 K^2 \right) F'(R) - 3 K \frac{d F'(R)}{d\tau}
+ \frac{\kappa^2}{a^6} \rho_\mathrm{matter}\, , \\
\label{Uni15}
0 = & \frac{a^{-6}}{2} \left( F(R) - 2 \kappa^2 \lambda \right)
 - \left( \dot K + 6 K^2 \right) F'(R)
+ 5 K \frac{d F'(R)}{d\tau} + \frac{d^2 F' (R)}{d\tau^2}
+ \frac{\kappa^2}{a^6} p_\mathrm{matter} \, ,
\end{align}
Here $K$ is defined by $K(\tau)=\frac{1}{a} \frac{da}{d\tau}$
and $\rho_\mathrm{matter}$ and $p_\mathrm{matter}$ stand for the energy-density and
pressure of the matter fluids present, respectively.
The ``prime'' and ``dot'' denote as usual differentiation with respect to
the Ricci scalar and with respect to $\tau$.
Equations (\ref{Uni14}) and (\ref{Uni15}) can be further combined to yield
\be
\label{Uni16}
0 = \left( 2 \dot K + 6 K^2 \right) F'(R)
+ 2 K \frac{d F'(R)}{d\tau} + \frac{d^2 F' (R)}{d\tau^2}
+ \frac{\kappa^2}{a^6} \left( \rho_\mathrm{matter} + p_\mathrm{matter}
\right)
\, .
\ee
Basically, the reconstruction method for the vacuum unimodular $F(R)$
gravity,
which we proposed in \cite{Nojiri:2015sfd} is based on Eq.~(\ref{Uni16}).
If we give the explicit form of $a=a\left(\tau\right) $ and therefore
$K\left(\tau\right)$, Eq.~(\ref{Uni16}) becomes a differential equation,
which is solved with respect to the function $F' = F'(\tau)$.
On the other hand, by using Eq.~(\ref{scalarunifrw}), one can obtain the
function $R=R(\tau)$, which can be solved with respect to $\tau$,
$\tau=\tau\left( R \right)$.
Substituting the form of $F' = F'(\tau)$, we obtain $F'$ as a function of
$R$, that is, $F'=F'(R)$.

\section{Newton's law in covariant unimodular $F(R)$ gravity}

Due to the unimodular constraint (\ref{UF1}), the unimodular gravity does
not have full covariance. The covariant formulation of the unimodular Einstein gravity has been
proposed
in Ref. \cite{Henneaux:1989zc}. By using such a formulation, we may start from the following action,
\be
\label{LUF1}
S = \int d^4 x \left\{ \sqrt{-g} \left( \mathcal{L}_\mathrm{gravity}
 - \lambda \right) + \lambda
\epsilon^{\mu\nu\rho\sigma} \partial_\mu a_{\nu\rho\sigma}
\right\}
+ S_\mathrm{matter} \left( g_{\mu\nu}, \Psi \right)\, ,
\ee
where $\mathcal{L}$ is the Lagrangian density of any gravity theory, and 
$a_{\nu\rho\sigma}$ is a three-form field.
By varying with respect to $a_{\nu\rho\sigma}$, gives the equation
$0 = \partial_\mu \lambda$, that is, $\lambda$ is a constant.
On the other hand, the variation with respect to $\lambda$ gives,
\be
\label{LLUF19}
\sqrt{-g} = \epsilon^{\mu\nu\rho\sigma} \partial_\mu a_{\nu\rho\sigma}\, ,
\ee
instead of the unimodular constraint. Because Eq.~(\ref{LLUF19}) can be solved with respect to $a_{\mu\nu\rho}$,
there is no constraint on the metric $g_{\mu\nu}$.

We now consider $a_{\mu\nu\rho}$, which has apparently four degrees
of freedom. The action is invariant under the gauge transformation
$\delta a_{\mu\nu\rho} = \partial_\mu b_{\nu\rho} + \partial_\nu b_{\rho\mu}
+ \partial_\rho b_{\mu\nu}$.
Here $b_{\mu\nu}$ is an anti-symmetric tensor field, that is,
$b_{\mu\nu} = - b_{\nu\mu}$, which has
apparently six degrees of freedom.
We should note that the gauge transformation is invariant under another
gauge transformation,
$\delta b_{\mu\nu} = \partial_\mu c_\nu - \partial_\nu c_\mu$.
Here $c_\mu$ is a vector field, which has four degrees of freedom.
Again we should note that the gauge transformation of the original gauge
transformation is invariant under the gauge transformation
$\delta c_\mu = \partial_\mu \varphi$.
The field $\varphi$ is a scalar field with one degree of freedom.
Therefore, the number of degrees of freedom in the gauge transformation
(\ref{LUF2}) is $6-4+1=3$ and
the number of degrees of freedom of $a_{\mu\nu\rho}$ is $4-3=1$.
Hence, one may choose the following gauge condition,
\be
\label{LUF5}
a_{tij} \left( = a_{jti} = a_{ijt} \right) = 0\, , \quad i,j=1,2,3\, .
\ee
Therefore the only remaining degree of freedom is given by $a_{ijk}$
$\left(  i,j,k = 1,2,3 \right)$.
Then we find,
\be
\label{LUF6}
S_{\lambda \alpha} = \int d^4x \lambda \left( - \sqrt{-g}
+ \epsilon^{\mu\nu\rho\sigma} \partial_\mu a_{\nu\rho\sigma} \right)
= \int d^4x \lambda \left( - \sqrt{-g}
+ \partial_t \alpha \right)\, ,
\ee
where $\alpha \equiv \frac{1}{3!} a_{123}$.
The system described by (\ref{LUF6}) might give a correction to Newton's
law, or it might be a ghost and generate negative norm states in the quantum
theory.
In order to consider the quantum system described by the action
(\ref{LUF6}),
we now investigate a similar quantum mechanical system.

We start with the system, where the Lagrangian is given by,
\be
\label{L1}
L= B y \dot x\, ,
\ee
which appears in the massless limit of the charged particle in uniform
magnetic
field $B$ with Lagrangian,
\be
\label{L2}
L= \frac{1}{2} m \left( {\dot x}^2 + {\dot y}^2 \right) + B y \dot x \, .
\ee
In (\ref{L1}) and (\ref{L2}), $B$ is a constant.
By using the Lagrangian (\ref{L1}), the equations of motion are given by
\be
\label{L3}
\dot x = \dot y = 0\, ,
\ee
that is, $x$ and $y$ are constant. On the other hand, the momenta $p_x$ and $p_y$ are given by
\be
\label{L4}
p_x \equiv \frac{\partial L}{\partial \dot x} = B y\, , \quad
p_y \equiv \frac{\partial L}{\partial \dot y}=0 \, ,
\ee
and therefore we obtain two constraints,
\be
\label{L5}
\chi_1 \equiv y - p_x=0\, , \quad \chi_2 \equiv p_y = 0 \, .
\ee
In order to consider the quantization of the system, we use Dirac's
formulation. First we introduce the matrix $C$ as follows,
\be
\label{L6}
C = \left( C_{ij} \right)
\equiv \left( \left[ \chi_i, \chi_j \right]_P \right)
= \left( \begin{array}{cc} 0 & B \\ -  B & 0 \end{array} \right)\, .
\ee
Here $\left[ A, B \right]_P$ for any physical quantities $A$ and $B$, is the
Poisson bracket defined by,
\be
\label{L7}
\left[ A, B \right]_P \equiv \frac{\partial A}{\partial x}
\frac{\partial B}{\partial p_x}
 - \frac{\partial A}{\partial p_x} \frac{\partial B}{\partial x}
+ \frac{\partial A}{\partial y} \frac{\partial B}{\partial p_y}
 - \frac{\partial A}{\partial p_y} \frac{\partial B}{\partial y} \, .
\ee
Then one can define the Dirac bracket as follows,
\be
\label{L8}
\left[ A, B \right]_D \equiv \left[ A, B \right]_P - \sum_{i,j=1,2}
\left[ A, \chi_i \right]_P C^{-1}_{ij}
\left[ \chi_j, B \right]_P\, .
\ee
Hence we find that,
\be
\label{L9}
\left[ x, p_x \right]_D = - \left[ p_x, x \right]_D= B \left[ x ,y \right]_D
= - B \left[y,x \right]_D = 1\, ,
\ee
and all the other Poisson brackets including $\left[y, p_y\right]_D$ vanish.
The quantization can be obtained by replacing the Dirac bracket with the
commutator and $1$ with $i$,
\be
\label{L10}
\left[ x, p_x \right] = - \left[ p_x, x \right]= B \left[ x ,y \right]
= - B \left[y,x \right] = i\, ,\quad \mbox{others}=0\, .
\ee
We also find that the Hamiltonian vanishes, that is,
\be
\label{L11}
H= \dot x p_x + \dot y p_y - L =0\, .
\ee
Therefore, there is no time evolution, which is consistent with the classical
solutions of Eq.~(\ref{L3}). The Hamiltonian (\ref{L11}) also shows that the states are infinitely
degenerate but there is no transition between the states and therefore, the
states are stable and there does not appear any ghost.

By a similar treatment, we find the following commutation relations for the fields
$\lambda$ and $\alpha$ similar to (\ref{L10}) in (\ref{LUF6}), which are,
\be
\label{LUF7}
\left[ \alpha, \lambda \right] = i \delta \left( \bm{x} \right)\, ,
\ee
with $\bm{x} = \left( x^1, x^2, x^3 \right)$. However, the Hamiltonian $H$ does not vanish, a result which is different from
(\ref{L11}), but $H$ is given by,
\be
\label{LUF8}
H = \int_S \mathrm{d} S \sqrt{-g} \lambda \, ,
\ee
where $S$ is an arbitrary space-like surface. Then although $\lambda$ is a constant, the time evolution of $\alpha$
is given by,
\be
\label{LUF9}
\frac{d\alpha}{dt} = i \left[ H, \alpha \right] =  \sqrt{-g}\, ,
\ee
which is consistent with the classical equation given by the variation of
the action (\ref{LUF6}) with respect to $\lambda$.
The eigenstate of the Hamiltonian $H$ could be given by the eigenstate of
$\lambda$. In the representation of the states by using $\alpha$, the commutation
relation
(\ref{LUF7}), yields $\lambda = i \frac{\delta}{\delta \alpha}$.
Then, the eigenstate $\Psi_{\lambda_0}(\alpha)$ of $\lambda$ with the
eigenvalue $\lambda_0$ is expressed as,
\be
\label{LUF11b}
\Psi_{\lambda_0}(\alpha) = \exp \left( i \lambda_0 \int_S \mathrm{d}S \alpha
\left( \bm{x} \right) \right)\, .
\ee
The eigenvalue of the Hamiltonian (\ref{LUF8}) is infinite, due to the
infinite volume of $S$ and furthermore unbounded from below. However, note that there is
no transition between the states and therefore the states are stable.

For the case of covariant unimodular $F(R)$ gravity, in which case,
\be
\label{LUF10}
\mathcal{L}_\mathrm{gravity} = \frac{F(R)}{2\kappa^2}\, ,
\ee
as in the standard $F(R)$ gravity, we may rewrite the action in the scalar-tensor form,
\be
\label{LUF11}
S = \int d^4 x \left\{ \sqrt{-g} \left( \frac{1}{2\kappa^2} \left( R
 - \frac{3}{2} g^{\mu\nu} \partial_\mu \phi \partial_\nu \phi - V(\phi)
\right)
 - \lambda \e^{2\phi} \right) + \lambda \epsilon^{\mu\nu\rho\sigma}
\partial_\mu a_{\nu\rho\sigma} \right\}
+ S_\mathrm{matter} \left( \e^\phi g_{\mu\nu}, \Psi \right)\, .
\ee
Here $V(\phi)$ is given by (\ref{UF4}) and the function $A(\phi)$ is defined by the
algebraic equation $\phi = - \ln F'(A)$. Note that $A=R$ in the original Jordan frame Lagrangian density
(\ref{LUF10}). In the action (\ref{LUF11}), one obtains $0 = \partial_\mu \lambda$ and
therefore $\lambda$ is a constant in this case too. Thus, the potential $V(\phi)$ is effectively changed as
\be
\label{LUF12}
V(\phi) \to \tilde V(\phi) = \frac{A(\phi)}{F'\left( A \left( \phi \right)
\right)}
 - \frac{F\left( A \left( \phi \right) \right)}{F'\left( A \left( \phi
\right)
\right)^2} + 2\kappa^2 \lambda \e^{2\phi}\, ,
\ee
Then if the mass of $\phi$, which is defined by
\be
\label{LUF13}
m_\phi^2 = \frac{3}{2} \frac{d^2 \tilde V(\phi)}{d\phi^2}\, ,
\ee
is small, there could appear a large correction to Newton's law of gravity.
By using the equation $\phi = - \ln F'(A)$ and (\ref{UF12}), the explicit
expression of $m_\phi^2$ is given by
\be
\label{ULF14}
m_\phi^2 = \frac{3}{2} \left\{
\frac{A(\phi)}{F'\left(A\left(\phi\right)\right)}
 - \frac{4F\left( A \left(\phi\right) \right) }{F' \left( A
\left(\phi\right)
\right)^2 }
+ \frac{1}{F'' \left( A \left(\phi\right) \right) }
+ \frac{8 \kappa^2 \lambda}{ F' \left( A \left(\phi\right) \right)^2 }
\right\}
\, .
\ee
The last term is a characteristic feature of the unimodular $F(R)$ gravity. As $\lambda$ is constant, the last term can be absorbed into the
redefinition
of $F(R)$, $F(R) \to F(R) + 2 \kappa^2 \lambda$.
The expression of $m_\phi^2$ obtained by the redefinition is identical with
the
expression in the standard $F(R)$ gravity.
Thus, there is no any essential difference in the corrections of Newton's law, 
between
the covariant unimodular $F(R)$ gravity  and the standard $F(R)$ gravity
\cite{Capozziello:2011et,Nojiri:2010wj,Nojiri:2006ri,Capozziello:2010zz}.
This is quite different from the non-covariant unimodular $F(R)$ gravity,
where standard Newton's law is recovered.

Let us study the FRW cosmology in the covariant unimodular gravity $F(R)$
((\ref{LUF1})
with (\ref{LUF10})).
Due to the absence of unimodular constraint (\ref{UF1}) in the case of the
covariant unimodular gravity, one may assume the standard FRW metric.
By the variation with respect to the metric we obtain,
\be
\label{JGRG13U}
0 =\frac{1}{2}g_{\mu\nu} \left(F(R) - 2\kappa^2 \lambda \right)
 - R_{\mu\nu} F'(R) - g_{\mu\nu} \Box F'(R)
+ \nabla_\mu \nabla_\nu F'(R) + \frac{\kappa^2}{2}T_{\mu\nu}\, .
\ee
and the following FRW equations are obtained:
\begin{align}
\label{JGRG15U}
0 =& -\frac{F(R)- 2\kappa^2 \lambda}{2} + 3\left(H^2 + \frac{dH}{dt} \right)
F'(R) - 18 \left( 4H^2 \frac{dH}{dt} + H \frac{d^2 H}{dt^2} \right) F''(R)
+ \kappa^2 \rho_\mathrm{matter}\, ,\\
\label{Cr4bU}
0 =& \frac{F(R) - 2\kappa^2 \lambda}{2} - \left( \frac{dH}{dt}
+  3H^2\right)F'(R) + 6 \left( 8H^2 \frac{dH}{dt}
+ 4 \left( \frac{dH}{dt} \right)^2 + 6 H \frac{d^2 H}{dt^2}
+ \frac{d^3 H}{dt^3} \right) F''(R) \nn
& + 36\left( 4H \frac{dH}{dt} + \frac{d^2 H}{dt^2} \right)^2 F'''(R)
+ \kappa^2 p_\mathrm{matter}\, .
\end{align}
Here, the Hubble rate $H$ is defined by $H=\frac{1}{a} \frac{da}{dt}$ as
usually, and the scalar curvature $R$ is given by $R=12H^2 + 6\frac{dH}{dt}$.
Rewriting Eqs.~(\ref{JGRG15U}) and (\ref{Cr4bU}) as follows,
\begin{align}
\label{CUF2}
0 =& -\frac{F(R)- 2\kappa^2 \lambda}{2} + 3\left(H^2 + \frac{dH}{dt} \right)
F'(R) - 3 H \frac{d F'(R)}{dt} + \kappa^2 \rho_\mathrm{matter}\, ,\\
\label{CUF3}
0 =& \frac{F(R) - 2 \kappa^2 \lambda}{2} - \left( 3H^2 + \frac{dH}{dt}
\right)F'(R) + 2 H \frac{d F'(R)}{dt} + \frac{d^2 F'(R)}{dt^2}
+ \kappa^2 p_\mathrm{matter}\, .
\end{align}
and by eliminating $\lambda$ from Eqs.~(\ref{CUF2}) and (\ref{CUF3}), we obtain
\be
\label{CUF4}
0 = 2 H \frac{dH}{dt} F'(R) - H \frac{d F'(R)}{dt} + \frac{d^2 F'(R)}{dt^2}
+ \kappa^2 \left( \rho_\mathrm{matter} + p_\mathrm{matter} \right)\, .
\ee
We should note that owing to the fact that $H = a^3 K$,
$\frac{dH}{dt} = 3 a^6 \dot K + a^6 K^2$,
Eq.~(\ref{CUF4}) in the covariant formulation is identical with
Eq.~(\ref{Uni16}) in the non-covariant formulation. In other words, we
proved
the dynamical equivalence of two formulations in the FRW background 
evolution level.
This is, of course, not accidental because Eq.~(\ref{JGRG13U}) in the
non-covariant formalism is identical with Eq.~(\ref{Uni12}).
The difference is that $\lambda$ is assumed to be a constant in
Eq.~(\ref{JGRG13U}) but $\lambda$ is not always invariant
in Eq.~(\ref{Uni12}).
Therefore the equations obtained by deleting $\lambda$ can be identical with
each other.
Hence, the background FRW cosmology in the covariant and the non-covariant
unimodular $F(R)$ gravity is just the same.
Nevertheless, due to possible corrections to Newton's law in covariant theory,
the Universe's structure formation may be different in the non-covariant and
covariant unimodular $F(R)$ gravity.

\section{Summary and Discussion}

In summary, we formulated the covariant unimodular $F(R)$ gravity and
demonstrated
that the resulting FRW cosmology is equivalent to the one corresponding to the non-covariant
version. We also demonstrated that Newton's law of gravity in the non-covariant formulation does not
change if compared with standard Einstein gravity result. At the same time,
Newton's law in the context of covariant unimodular $F(R)$ gravity is modified just in
the same way as in the ordinary $F(R)$ gravity theory
\cite{Capozziello:2011et,Nojiri:2010wj,Nojiri:2006ri,Capozziello:2010zz}.

One may consider some extension of the covariant formulation.
The unimodular gravity was proposed in order to solve the problem of the
large vacuum energy. In the unimodular gravity, the cosmological constant $\Lambda$ which
includes the large vacuum energy, can be absorbed into the constant shift of the
Lagrange multiplier field $\lambda$ as follows $\lambda \to \lambda - \Lambda$.
Therefore, the cosmological constant $\Lambda$ does not affect the dynamics.

If we consider models which have similar properties, the problem of the
vacuum energy could be solved. In \cite{Nojiri:2016mlb}, a new extension of the covariant unimodular
gravity has been proposed.
The action of the model is given by
\be
\label{CCC2}
S = \int d^4 x \sqrt{-g} \left\{ \mathcal{L}_\mathrm{gravity}
 - \lambda \left( 1 - \frac{1}{\mu^4}\nabla_\mu J^\mu  \right) \right\}
+ S_\mathrm{matter} \, ,
\ee
where $\mu$ is a constant with mass dimensions, $J^\mu$ is a general vector
quantity, and finally $\nabla_\mu$ is a covariant derivative with respect to the
vector field. Dividing the gravity Lagrangian density $\mathcal{L}_\mathrm{gravity}$ into
the
sum of the cosmological constant $\Lambda$ and another part
$\mathcal{L}_\mathrm{gravity}^{(0)}$ and redefining the Lagrange multiplier field $\lambda$
by
$\lambda \to \lambda - \Lambda$, we can rewrite the action (\ref{CCC2}) as
follows,
\be
\label{CCC2b}
S = \int d^4 x \sqrt{-g} \left\{ \mathcal{L}_\mathrm{gravity}^{(0)}
 - \lambda \left( 1 - \frac{1}{\mu^4}\nabla_\mu J^\mu  \right) \right\}
+ S_\mathrm{matter}
 -  \frac{\Lambda}{\mu^4} \int d^4 x \sqrt{-g} \nabla_\mu J^\mu \, .
\ee
Because the integrand in the last term is a total derivative, the last term
does
not affect the dynamics and one may drop the last term, again.
In Ref.~\cite{Nojiri:2016mlb}, a model using the topological field theory
(\cite{Witten:1988ze}) was studied. Here we may propose a new class of models, whose action is given by
\be
\label{NewCCC}
S = \int d^4 x \sqrt{-g} \left\{ \mathcal{L}_\mathrm{gravity}
 - \lambda  + \mathcal{L}_\lambda \left( \partial_\mu, \partial_\mu \lambda,
\varphi_i \right) \right\}
+ S_\mathrm{matter} \, ,
\ee
where $\mathcal{L}_\lambda \left( \partial_\mu, \partial_\mu \lambda,
\varphi_i
\right)$ is the Lagrangian density including the derivatives of $\lambda$
and other
fields $\varphi_i$, but not including $\lambda$ without derivative.
Hence, if we divide the  Lagrangian density $\mathcal{L}_\mathrm{gravity}$
into
the sum of the cosmological constant $\Lambda$ and other part
$\mathcal{L}_\mathrm{gravity}^{(0)}$, the cosmological constant can be
absorbed into the redefinition of the Lagrange multiplier field
$\lambda$, $\lambda \to \lambda - \Lambda$.

\section*{Acknowledgments.}

This work is supported by MINECO (Spain), project FIS2013-44881 and I-LINK
1019, by JSPS fellowship
ID No.:S15127 (SDO) and (in part) by MEXT
KAKENHI Grant-in-Aid for Scientific Research on Innovative Areas
  (No. 15H05890) and the
JSPS Grant-in-Aid for Scientific Research (C) \# 23540296 (S.N.)


\begin{thebibliography}{99}

\bibitem{Alvarez:2005iy}
E.~Alvarez,
JHEP {\bf 0503} (2005) 002
doi:10.1088/1126-6708/2005/03/002
[hep-th/0501146].

\bibitem{Nojiri:2015sfd}
S.~Nojiri, S.~D.~Odintsov and V.~K.~Oikonomou,
arXiv:1512.07223 [gr-qc].

\bibitem{Capozziello:2011et}
S.~Capozziello and M.~De Laurentis,
Phys.\ Rept.\  {\bf 509} (2011) 167
doi:10.1016/j.physrep.2011.09.003
[arXiv:1108.6266 [gr-qc]].

\bibitem{Nojiri:2010wj}
S.~Nojiri and S.~D.~Odintsov,
Phys.\ Rept.\  {\bf 505} (2011) 59
doi:10.1016/j.physrep.2011.04.001
[arXiv:1011.0544 [gr-qc]].

\bibitem{Nojiri:2006ri}
S.~Nojiri and S.~D.~Odintsov,
eConf C {\bf 0602061} (2006) 06
[Int.\ J.\ Geom.\ Meth.\ Mod.\ Phys.\  {\bf 4} (2007) 115]
doi:10.1142/S0219887807001928
[hep-th/0601213].

\bibitem{Capozziello:2010zz}
V.~Faraoni and S.~Capozziello,
Fundam.\ Theor.\ Phys.\  {\bf 170} (2010).
doi:10.1007/978-94-007-0165-6

\bibitem{Nojiri:2016ygo}
S.~Nojiri, S.~D.~Odintsov and V.~K.~Oikonomou,
arXiv:1601.04112 [gr-qc].

\bibitem{Saez-Gomez:2016gum}
D.~Saez-Gomez,
arXiv:1602.04771 [gr-qc].

\bibitem{Lim:2010yk}
E.~A.~Lim, I.~Sawicki and A.~Vikman,
JCAP {\bf 1005} (2010) 012
doi:10.1088/1475-7516/2010/05/012
[arXiv:1003.5751 [astro-ph.CO]].

\bibitem{Capozziello:2010uv}
S.~Capozziello, J.~Matsumoto, S.~Nojiri and S.~D.~Odintsov,
Phys.\ Lett.\ B {\bf 693} (2010) 198
doi:10.1016/j.physletb.2010.08.030
[arXiv:1004.3691 [hep-th]].

\bibitem{Henneaux:1989zc}
M.~Henneaux and C.~Teitelboim,
Phys.\ Lett.\ B {\bf 222} (1989) 195.
doi:10.1016/0370-2693(89)91251-3

\bibitem{Nojiri:2016mlb}
S.~Nojiri,
arXiv:1601.02203 [hep-th].

\bibitem{Witten:1988ze}
E.~Witten,
Commun.\ Math.\ Phys.\  {\bf 117} (1988) 353.
doi:10.1007/BF01223371

\end{thebibliography}
\end{document}